# In Situ Growth of Copper Channels within CuCl and PVDF Composite for Durable WORM Device Formation[*]

Shilpi Bose,[†] Aloka Sinha,[‡] and Santanu Ghosh[§]

*Indian Institute of Technology, Delhi*

(Dated: July 30, 2024)

This study details the creation of a Write Once-Read Many (WORM) memory device utilizing cuprous chloride (CuCl) and poly(vinylidene fluoride-co-hexafluoropropylene) (PVDF-HFP) polymers. Employing an in-plane configuration, a deliberate 1:10 ratio of CuCl to PVDF-HFP has been selected. This ratio aims to establish an in-situ copper channel within the device. The electrical response exhibits consistent memory retention over an extended duration. The WORM characteristics are attributed to the development of multiple conducting filaments or a highly conductive percolative path created by Cu ions within the polymer matrix. The UV-Vis study also reinforces the obtained results. Additionally, the memristor undergoes specific poling and cooling conditions. The fabrication approach employed in this research yields a distinctive type of Resistive Switching Device (RSD). Once the device is activated, it maintains its state even after the applied field is reduced. This specialized device holds potential applications in Compact Disc-Recordable (CD-R), Digital Versatile Disc-Recordable (DVD-R), Blu-ray Disc Recordable (BD-R), and Write Once USB Drives.

## I. INTRODUCTION

A resistive switching device (RSD), commonly referred to as the ideal memristor, is a unique device that exhibits a variation in resistance when subjected to a specific positive current, which remains constant until a 'reset' negative current is applied along with a negative voltage[1, 2]. As a result, the device can be characterized by its two distinct resistance states: SET and RESET [3, 4].

Current literature focuses on studying and documenting RSDs fabricated using inorganic materials, which are frequently reported due to their well-controlled structure, ease of deposition on thin films, and strong compatibility with complementary metal oxide semiconductor (CMOS) processes[5, 6]. The prevalent explanation for the resistive switching mechanism, in this case, is the movement of oxygen vacancies driven by an applied electric field [7, 8].

Organic RSDs represent a distinct category with an ongoing debate surrounding their operational mechanism [9, 10]. However, polymer materials within this category are considered highly promising for the advancement of RSDs due to their affordability, ease of processing, mechanical flexibility, and the ability to finely tune their electronic properties through innovative molecular design. They may also be made using nanocomposite (NC) materials, which involve blending an insulating polymer matrix with active fillers or nanoparticles (NPs)[11–15]. This combination allows for enhanced functionality and performance in RSDs [16–18].

There are two distinct approaches to producing nanocomposite materials. The first approach involves directly adding nanometric fillers into the polymer matrix (ex-situ method). Alternatively, in the second approach, precursor materials of the nanophase are mixed into the polymer matrix, and the NPs are formed in-situ during polymerization or through a dedicated post-processing step [16].

A Write Once Read Many (WORM) RSD device exhibits a non-volatile memory behavior [19]. It allows for data to be written once to the device and read multiple times thereafter. Once data is written to specific memory cells in the device, it becomes permanently stored and cannot be modified or erased. WORM RSDs are commonly used in applications where data integrity and non-alterability are critical, such as in archival storage systems or secure data recording.

Organic write once read many (oWORM) devices offers a multitude of advantages that contribute to its appeal in the realm of electronic systems. Firstly, these devices exhibit a commendable level of energy efficiency, a critical factor in contemporary electronics, making them suitable for battery-operated devices and energy-conscious applications. Their unique property of being compatible with flexible substrates enables the creation of electronics that are not only robust but also adaptable to unconventional form factors. This attribute finds utility in wearable technology, flexible displays, and conformal electronics, expanding the horizons of device design [20].

Moreover, oWORM devices are characterized by their ease of fabrication, often relying on techniques such as inkjet printing and roll-to-roll processing [16, 21]. This simplicity in manufacturing can potentially lead to reduced production costs, which is a significant consideration in today's competitive electronics market. Their compatibility with large-area manufac-turing further distinguishes them, allowing for the creation of expansive surfaces with integrated functionalities, thereby finding applications in areas like large-scale displays and sensor arrays. The non-volatile memory capabilities of oWORM devices ensure data retention even in the absence of power, a vital attribute for applications requiring persistent storage of information, such as configuration data. Additionally, the limited data alteration property enhances security, making these devices suitable for tamper-proof data storage, cryptographic key management, and secure boot mechanisms[22].

---

[*] A footnote to the article title
[†] sbshilpibose@gmail.com
[‡] Aloka.Sinha@physics.iitd.ac.in
[§] santanu1@physics.iitd.ac.in

From an environmental perspective, organic materials used in these devices often have a lower environmental impact compared to traditional inorganic semiconductors, aligning with the increasing demand for eco-friendly technologies. Furthermore, the innovative potential of oWORM devices is noteworthy, as their distinct characteristics enable new and unconventional applications that might be challenging to achieve using conventional materials[23].

However, it is important to acknowledge that oWORM devices also face challenges. These include ensuring stability and reliability over time, addressing potential performance limitations, and advancing manufacturing processes to meet the demands of complex applications. Balancing these benefits and challenges is crucial for the successful integration of oWORM devices in the broader landscape of electronic technologies[24].

In this paper, an in-situ method is employed to prepare a WORM device formed by NC of CuCl and PVDF-HFP. This method is chosen as it is both straightforward and efficient in preventing particle agglomeration while ensuring a favorable spatial distribution within the polymer matrix. This method also eliminates the need for specialized expertise and training required for designing new polymers and handling them. PVDF-HFP is selected as the polymer matrix for such applications due to its notable attributes, including a high dielectric constant, chemical stability, and robust mechanical strength. This polymer serves as a quasi-solid medium, effectively facilitating the movement of metal ions of copper and supporting the overall ionic movements within the device when an electric field is applied. Our endeavor aims to strike a balance between the pros and cons inherent in utilizing a commercial PVDF polymer as the foundation for an oWORM device, with a view toward its potential applications in the future. The fabrication technique employed for the device is relatively straightforward.

### A. Experimental Details

#### 1. Materials and Composite preparation

CuCl and PVDF-HFP are purchased from Sigma Aldrich and used without further purification. The preparation of the resistive switching material, the NC polymer, involves several steps. First, a solution of PVDF-HFP in DMF (1gm:10ml) is obtained by vigorously stirring the mixture until the complete dissolution of the polymer in the solvent. Then, 50mg of CuCl is added to 5gm of the solution. The mixture is carefully mixed to ensure thorough dispersion of the latter within the polymer solution. This sequential process allows for the creation of the NC polymer, which exhibits WORM-like properties. The final product demonstrates a uniform distribution and consistency across its entirety, displaying a high degree of homogeneity. The same is seen in Fig. 1(a).

#### 2. Device fabrication

In the series of experiments performed, a chip utilizing a silicon substrate as its core foundation is utilized. This substrate served as the underpinning for the deposition of intricate gold patterns. The formation of the electrode design is achieved through the utilization of maskless lithography methods, succeeded by the deposition of a layer of titanium measuring 10 nm in thickness. This is further overlaid with a 100 nm thick layer of gold. The role of the titanium layer is to function as an adhesive agent, effectively reducing the lattice mismatch that exists between the silicon wafer and the gold layer. The excess metal is then removed using the lift-off technique, leaving behind the intended electrode design. In Fig. 1(b), the optical image of the channel within the fabricated device is illustrated. The electrodes are positioned with a gap of 10 $\mu$m between them, on top of which the active nanocomposite material is spin-coated at a rate of 3000 rpm for 20 sec[25]. The sample is left to dry at room temperature for several hours after that. Atomic force microscopy (AFM) is employed for conducting a comprehensive morphological analysis of both the fabricated chip and the NC samples. The morphology of the active matrix on the fabricated device is displayed in Fig. 1(c) and (d), while Fig. 1(e) validates the height of the deposited gold and titanium film. Fig. 1(f) shows the final fabricated device.

#### 3. FTIR of PVDF/HFP and CuCl compound

The PVDF-HFP matrix was verified through FTIR analysis, with the results presented in Fig. 2. Characteristic peaks of pure PVDF-HFP are observed at 836, 876, 1067, 1170, 1232, 1404, 2900, and 2989 $cm^{-1}$ [26]. The incorporation of CuCl salt into the PVDF-HFP matrix resulted in the emergence of new peaks, highlighted within a blue box in the figure, indicating the presence of the dissolved compound in the PVDF-HFP matrix.

tor under the influence of a strong electric field. The potential experienced by the electrons is triangular-shaped with curved edges[30]. Schottky transport is when an escaping electron in the metal interface after obtaining sufficient thermal energy to overcome a square barrier height of the metal-dielectric interface, flows and results in the conduction in the system[31]. The logarithm of current (log I) when differentiated with the logarithm of voltage V, results in a constant alpha, which when plotted against the square root of potential and subjected to linear fitting, yields the relative permittivity of the sample($\epsilon_r$). However, after substituting and yielding the unrealistic values of $\epsilon_r$ in our case, this conduction mechanism is also ruled out and not further explored[32]. In Ohmic conductors, the resistance remains constant over a wide range of applied voltage. In ferroelectric insulators like PVDF-HFP, space charge current refers to the electric current resulting from the movement of free charges within the material due to the presence of a space charge. Current due to this effect has a quadratic relation with voltage[33]. A plot with a slope of 2 in log(I) versus log(V) indicates the presence of the aforementioned effect. Trap-charge limited current (TCLC) mechanism is when relation $I \propto V^\alpha$, where $\alpha$ represents a positive constant with values other than 1 or 2. $\alpha$ less and greater than 2, suggests an energetically uniform (continuous) or exponential distribution of traps within the system respectively[34, 35]. It means that free charges are trapped. The transition from SCLC to TCLC often occurs as the electric field intensifies. At lower fields, the space charge dominates, and carriers move relatively freely. However, as the field increases, carriers experience stronger forces and are more likely to interact with traps. This interaction becomes more prominent, leading to a transition from SCLC to TCLC behavior. In a poled sample it should be noted that, maximum fluorine atoms are oriented towards one electrode, while hydrogen atoms are directed towards the opposite electrode. This arrangement may be contributing to the formation of a space-charged layer within the sample. This layer potentially contains many Coulombic traps, curbing the migration of conducting carriers towards the electrodes and concurrently limiting the current [36]. Trap-assisted tunneling is a quantum mechanical phenomenon in which mobile charges such as electrons, overcome energy barriers by tunneling through traps within a device [18, 29, 34, 37].

### 2. I-V characteristics of fabricated device under ambient conditions

Initially, the electrical characterization of the chips is carried out using a Keithley 4200-SCS Semiconductor Characterization System. The characterization process involved utilizing a standard 2-point setup and establishing contact with the samples through tungsten micro-needles that are firmly placed directly on the gold electrodes. Voltage versus current hysteresis measurements are performed over a range of -10 volts to +10 volts, with a constant compliance current of 20 µA. The electrical characterization is conducted at room temperature.

A cyclic I-V (current vs. voltage) graph is plotted for the fabricated device. Initially, a negative potential of -8.8 V reveals a SET voltage. This voltage induces a transition in the device from a state of high resistance (HRS) to a state of low resistance (LRS). This change is linked to the development of a conductive pathway within the polymer matrix, enabled by copper ions through a percolation effect[9]. The copper ions originate from the ionization of copper chloride. The formation of both individual and multiple filaments of conducting copper within the NC is feasible. Subsequently, six cycles of current-voltage (I/V) testing are conducted in parallel. Remarkably, throughout these cycles, the device remains in its low resistance state, displaying notable stability by not reverting back to its original high resistance state. The obtained result is depicted in Fig. 3(a). This illustrates the device's WORM-like characteristics, where once information is stored, it becomes resistant to tampering. This could potentially be due to the filament's thickness or its branching into multiple paths that lead to a stable conducting pathway [16]. As the endurance test evaluates a memristor's ability to maintain functionality through repeated cycles, and the retention test assesses its capacity to uphold resistance state over time without active programming, both offer crucial insights into reliability and stability across applications. To gauge retention and endurance, we subjected the same device to numerous repetitive cycles and long-term testing. The retention plot is provided in Fig. 3(b). Throughout the testing period, the device exhibited a consistent conduction state, maintaining stability in both the low-resistance and high-resistance. This is evident from Fig. 3(e). The switching ratio, defined as the RESET/SET voltages, could not be determined for this device because it is a worm device, and the RESET voltage is not defined for such devices.

To elucidate the conduction mechanisms within the memrister at any stage, different methodologies have been employed. Under ambient conditions, the memristor undergoes initial reverse biasing. In the investigation of charge transport phenomena in the initial negative voltage sweep, a logI-logV graph is constructed and subsequently fitted with a linear regression. The same is plotted in Fig. 3(c). The slope of this line is 0.8, indicating predominantly Ohmic/drift conduction. Ohmic contact is prevalent when the energy levels of the polymer and metal are compatible to facilitate the movement of electrons without significant barriers. Mathematically, it is expressed by a linear relationship between voltage and current. Given that the work functions of Cu ($\phi_{Cu}$ = 4.65 eV) and Au ($\phi_{Au}$ = 4.8 eV) are lower than that of PVDF ($\phi_{PVDF}$ = 5.65 eV). Additionally, PVDF has an electron affinity ($\chi_{PVDF}$) of 3.9 eV and a bandgap (Eg) of 3.5 eV, which suggests ohmic interfaces [38]. It's also plausible that within the system, there exist some Coulombic barriers through which the conducting particles are making their transitions. After the SET state, Ohmic conduction from the metallic junction controls the carrier transport in the LRS. The graph in Fig. 3(d) illustrates a straight-line slope of approximately 1 in the logI-logV plot, which bolsters the prior argument.

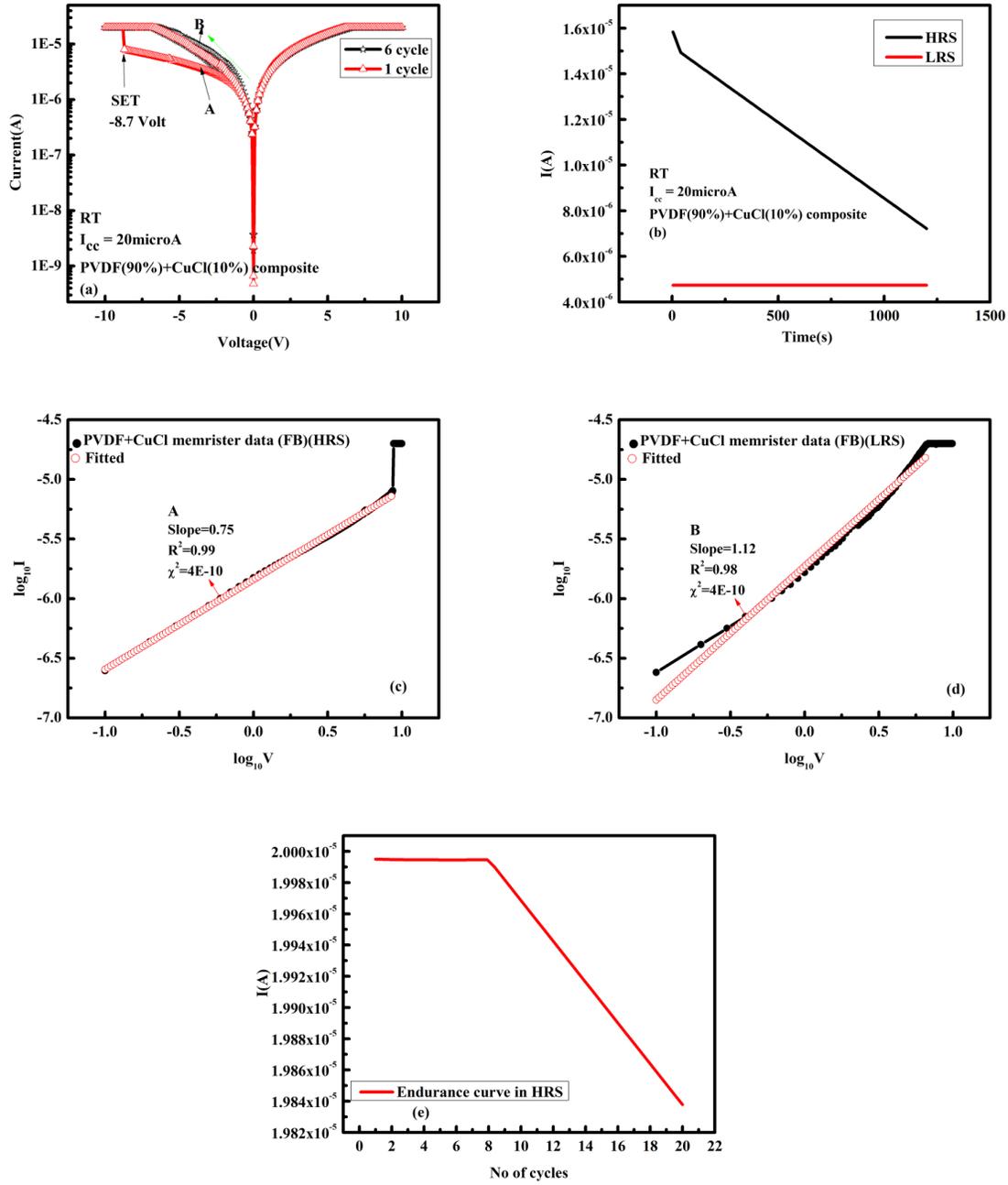

FIG. 3. (a) Cyclic current vs. voltage measurement (b)Retention plot of CuCl/PVDF – HFP nanocomposite chip (c) Conduction of CuCl/PVDF – HFP nanocomposite chip in the HRS under forward bias, in the absence of electric stressing condition (d) Conduction of CuCl/PVDF – HFP nanocomposite chip in the LRS under forward bias, in the absence of electric stressing conditions and (e) Endurance plot of CuCl/PVDF – HFP nanocomposite chip

### 3. UV-VIS characteristics of fabricated device

The UV-VIS spectrum of the pristine device as shown in Fig. 4 exhibits distinct peaks associated with $Z_3$ and $Z_{1,2}$ free excitons. The $Z_3$ free exciton peak appears at 381 nm (3.254 eV at room temperature), while the higher energy shoulder is attributed to the $Z_{1,2}$ peak. The band structure of CuCl diverges from the typical semiconductor arrangement, featuring the split-off hole ($\gamma_7$) positioned at the valence band's apex, around 60 meV distant from the degenerate heavy-hole and light-hole bands ($\gamma_8$). Traditionally, the exciton created by the electron at $\gamma_6$ and the hole at $\gamma_7$ or $\gamma_8$ is referred to as the $Z_3$ or $Z_{1,2}$ exciton [39], respectively. The reduction in this peak intensity after 10 measurement cycles indicates the formation of a copper channel within the system.

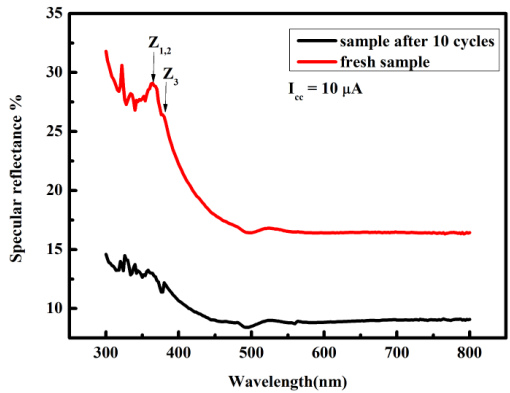

FIG. 4. (a) UV-VIS measurement in the pristine sample

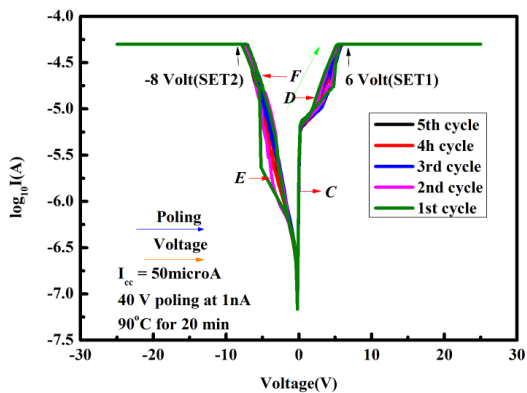

FIG. 5. Cyclic current vs. voltage measurement after sample poling and initial voltage sweep in the same direction.

*4. I-V characteristics of fabricated device under poling conditions*

Next, the sample is heated to 90$^o$C and subjected to a 40-voltage potential applied across an electrode gap of 10 microns for 20 minutes. The compliance is set as 1nA. The intentional use of such low compliance is aimed at averting the development of a conducting filament in the device during the poling phase. Afterward, the system is rapidly cooled to room temperature, and an I-V measurement from -25V to +25V is conducted in the presence of a compliance current limit of 50 $\mu$A. The biasing during both poling and measurements is maintained consistently. Fig. 5 illustrates the cyclic I-V characteristics of the PVDF-HFP and CuCl-based memristor under the aforementioned conditions. The compliance current in Fig. 3(a) and Fig. 5 has been elevated to facilitate the formation of the copper channel more effectively. At lower compliance levels during poling, the worm characteristics were not observed. This is due to the hindrance of effective copper atom migration towards the electrode caused by the formation of traps and space charge regions within the poled PVDF/HFP matrix, thereby preventing the formation of a conductive percolative path.

A forward sweep is carried out originally and the graph discloses two noteworthy regions, denoted as $C(i)$ and $C(ii)$. In the zone labeled as $C(i)$, the logI-logV graph (Fig. 6(a)) exhibits a linear fit with a slope equal to 0.4, signifying a continuous trap-charge limited current conduction characterized by a considerable density of trapped carriers within a specific, low voltage range. To scrutinize the conduction in region $C(ii)$, the I-V graph in this part is fitted with a linear line. The fit is significantly deviating from the observed data. It dictates that some other complex charge transport mechanisms may be occurring, particularly related to ion hopping[36, 40]. The shift from a linear to a curved section may possibly be due to ions and could provide valuable insights into their complex movement within the system. After this, a region of copper channel is formed within the host matrix and is characterized by SET1. Within the low-resistance state under forward bias, the regions of interest, as indicated in Fig. 6(b), are named as $D(i)$ and $D(ii)$. In the former and latter parts, the carriers i.e. electrons and holes become exponentially and continuously trapped within defect barriers respectively. A curvilinear sec-tion indicates the movement of ions away from the cathode in a complex fashion. This suggests that, during the system's presence in this low-resistance state with decreasing voltage in forward bias, the filament undergoes a gradual disintegration, resulting in the rupture of the copper channel. The conduction in reverse bias in HRS is further grouped into two regions, namely $E(i)$ and $E(ii)$ as depicted in (Fig. 6(c)). The logI-logV graph in part $E(i)$ indicates a slope of 1.7. It suggests a space-charge limited current associated with a sparse amount of trapped carriers within the system. The slope of 0.8 in Re-gion $E(ii)$ rather implies that the effect of poling is getting minuted in the presence of opposite polarity and the impact of trapping is becoming less profound. There is yet some divergence from the pure line feature here, which may be due to the conduction of ions. The low-resistance state (LRS) during reverse bias is denoted as Region $F$, and this region is subdivided into three domains: $F(i)$, $F(ii)$, and $F(iii)$. It is demonstrated in (Fig. 6(d)). In Region $F(i)$, charged particles experience exponential restriction back to the potential traps. Region $F(ii)$ corresponds to the part where the space-charge effect is minimal. Leakage current due to few ions also seems possible in the nonlinear region between $\boldsymbol{F(i)}$ and $\boldsymbol{F(ii)}$. Region $\boldsymbol{F(iii)}$ signifies the near space-charge formation within the samples. For each subsequent cycle, the filament strengthens progressively, becoming more robust with each iteration.

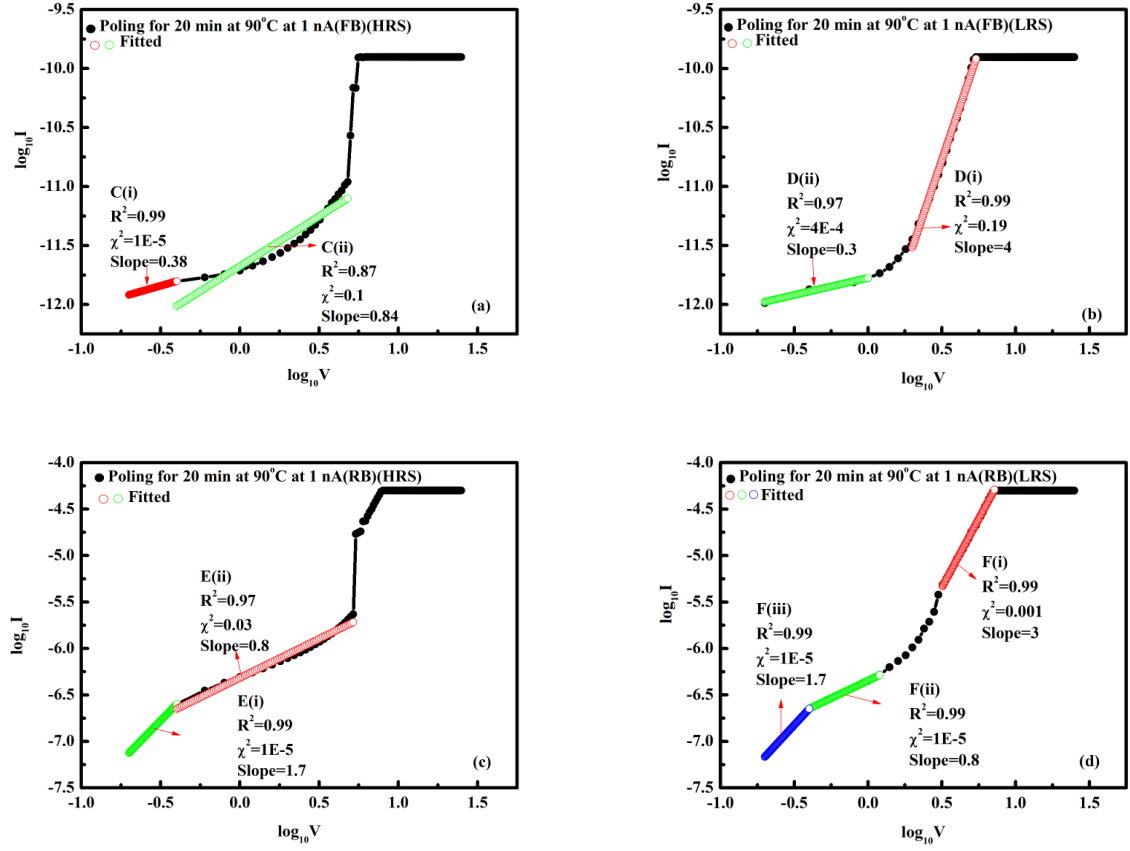

FIG. 6. (a) Mechanism of conduction after sample poling and initial voltage sweep in the same direction (positive bias).

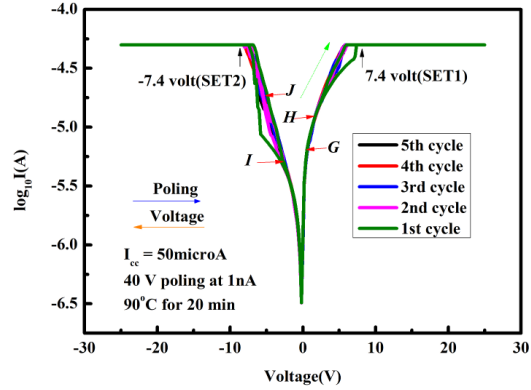

FIG. 7. (a) Cyclic current vs. voltage measurement after sample poling and initial voltage bias in the opposite direction.

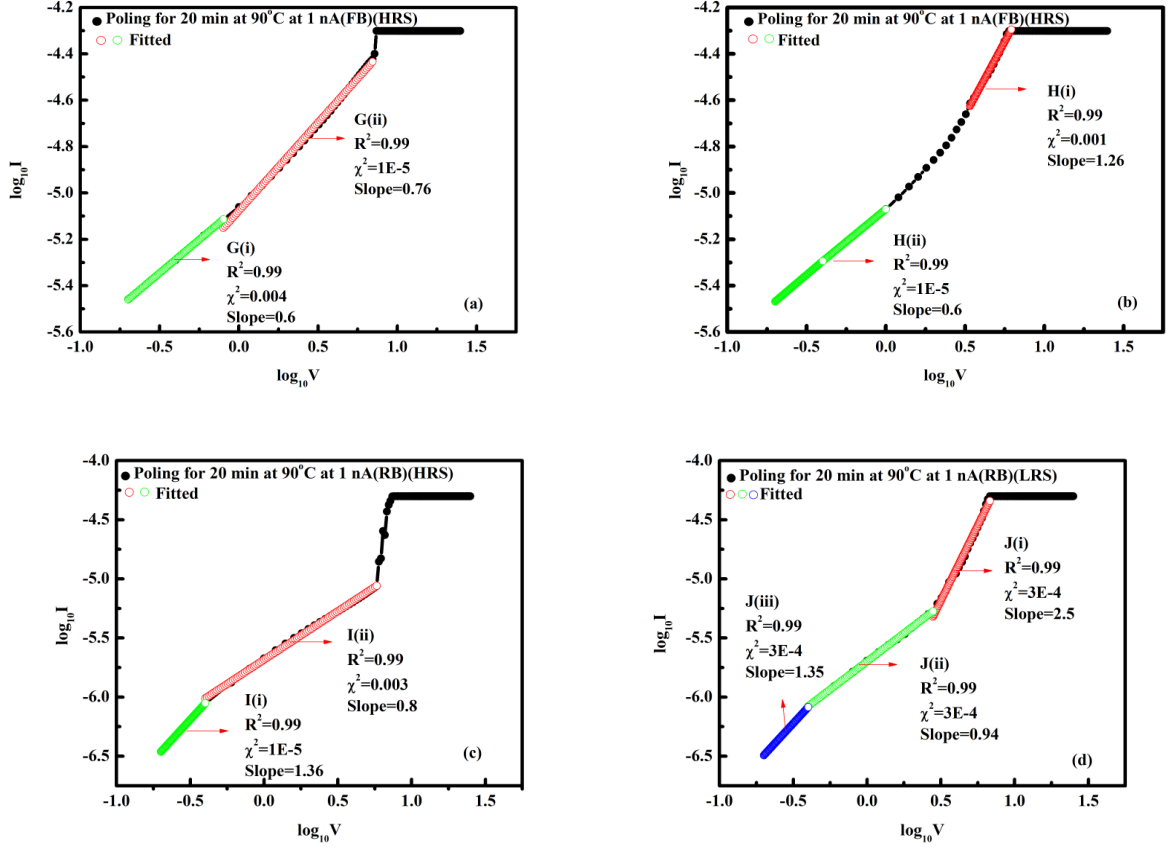

FIG. 8. (a) Mechanism of conduction after sample poling and initial voltage sweep in the opposite direction (positive bias).

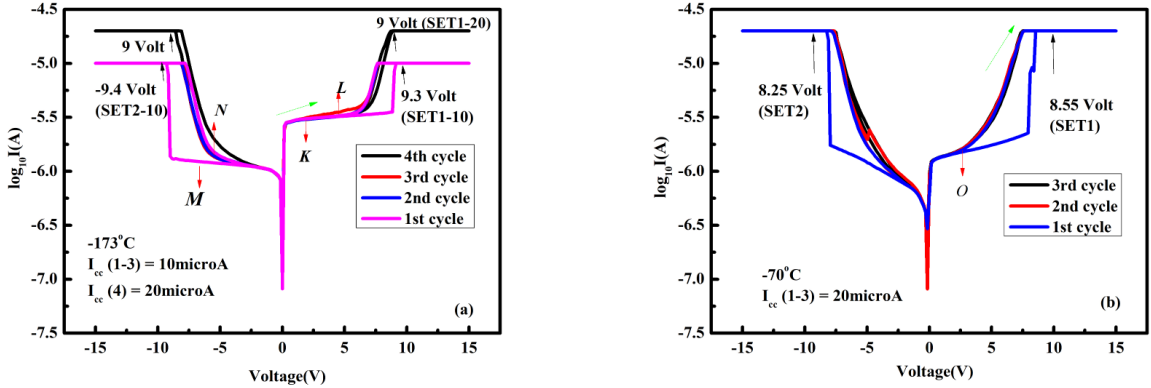

FIG. 9. (a) Cyclic current vs. voltage measurement after sample cooling.

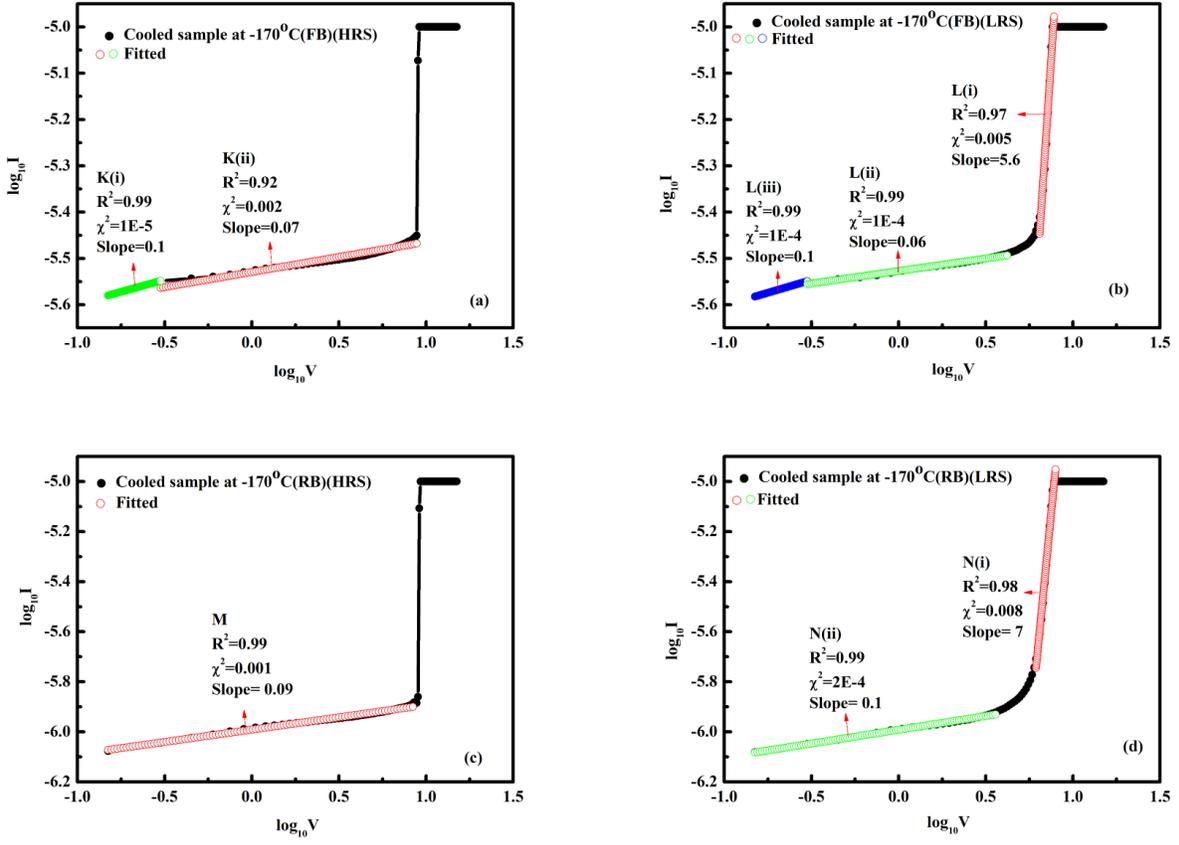

FIG. 10. (a) Mechanism of conduction after cooling to -170°C in forward bias.

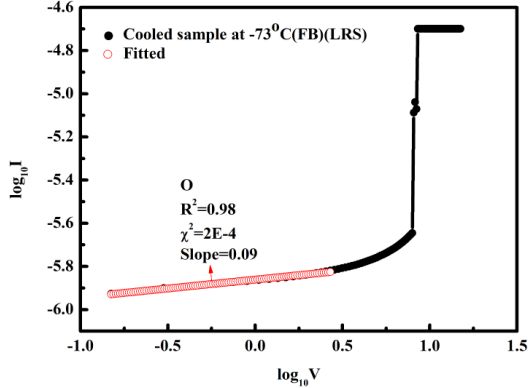

FIG. 11. (a) Mechanism of conduction after cooling to -73°C in forward bias.

Given the distinct variations observed in the initial sweep under both forward and reverse bias in Fig. 5, further investigation becomes imperative. Notably, a discernible shift in the SET voltage is evident during the reverse bias at -8V, denoted as SET2. The asymmetry between SET1 and SET2 voltages suggests potential challenges faced by transport carriers moving in a specific direction. This asymmetry could be attributed to obstacles encountered by carriers and the width of the filament atop the electrode. It is conceivable that the formation of a thinner filament during forward bias precedes the development of a thicker one during reverse bias. An additional SET voltage reinforces the instability of the channel established during the positive sweep, leading to rapid and spontaneous disintegration. Subsequent cycles result in filament branching, after which the memristor consistently maintains a state of low resistance (LRS) for a significant duration.

For the next experiment, another adjacent sample on the same chip underwent poling at $90^o$C for 20 minutes, maintaining a 1 nA compliance current. Subsequently, the sample was rapidly returned to ambient conditions. Notably, the polarity during poling and the acquisition of I-V (from -25V to +25V) characteristics are reversed this time. It was observed that the bias values for SET1 and SET2 exhibited nearly symmetric behavior from Fig. 7. This may occur due to the comparable movement of charged particles in both polarities. The voltage associated with SET2 could signify the branching of the filament into multiple directions. We again categorize conduction in forward bias under HRS into two regions, namely $\boldsymbol{G(i)}$ and $\boldsymbol{G(ii)}$. It is evidenced in (Fig. 8(a)). In these regions, we observe drift current and TCLC (having a uniform and moderate number of traps) as primary conduction mechanisms. Since the slopes are less than 1, it elucidates that there is the absence of a polar phase in the system (Fig. 8(a)). A slight shift from a rectilinear trait indicates the onset of complex ionic transport. After this, a stable pathway between the electrodes is formed by copper carriers. The LRS in forward bias indicates the formation of a channel with an average amount of uniform trapped current carrier concentration. A minuscule dissociation of copper ions from the bridge is also evident here as seen from Fig. 8(b). During the negative sweep, the I-V characteristics recorded in the high resistive state that there are some energetically continuous traps for the charges within the system. However, the primary conduction from the metallic bridge is Ohmic. This observation is supported by the characteristics of segments $\boldsymbol{I(i)}$ and $\boldsymbol{I(ii)}$, along with their associated slopes, as illustrated in Fig. 8(c). During the retraction of the negative sweep in the Low Resistive State (LRS), region $\boldsymbol{J(i)}$ is identified as a TCLC effect with exponential trapping of electrical entities with decreasing voltage. Subsequently, regions $\boldsymbol{J(ii)}$ and $\boldsymbol{J(iii)}$ suggest Ohmic behavior due to conduction from the branched filament. The same is unveiled from Fig. 8(d)

### 5. I-V characteristics of fabricated device under cooling conditions

The literature consistently highlights the significant influence of temperature on the concentration of trapped carriers within the metal-insulator-metal junction [34]. This phenomenon underscores the importance of understanding and controlling temperature conditions in order to modulate the concentration of trapped carriers effectively. Consequently, the samples underwent two consecutive temperatures (-170 and -$73^o$C), both below the glass transition temperature, and the observations are noted in (Fig. 9). The voltage is swept between -15V to +15V keeping $10\mu$A as the compliance current. The SET voltage was relatively elevated in the cooled samples, primarily because the lower temperature imposed restrictions on the motion of ions and electrons. It is observable that with the increase in compliance from 10 $\mu$A to 20 $\mu$A, the SET voltage decreased. This phenomenon might be attributed to the fact that the higher compliance enabled a greater flow of carriers, thereby reducing the SET voltage.

During the initial forward bias, the cyclic I-V curve at -$170^o$C is documented in (Fig. 10(a)) and segmented into two sectors, namely $\boldsymbol{K(i)}$ and $\boldsymbol{K(ii)}$. These portions display logI-logV curves with extremely small slopes, signifying the existence of numerous uniform and ordered traps in the system. This is a consequence of the limited energy available to electrons or ions to hop between these energy pits, resulting in the majority being trapped. In the HRS state under positive bias, a shift from region $\boldsymbol{L(i)}$ to $\boldsymbol{L(ii)}$ and subsequently $\boldsymbol{L(iii)}$ is observed. This is shown in (Fig. 10(b)). It implies a homogenization and an increase in the number of sites where carriers are captured. The cusped region is suggestive of multiple transitions, particularly associated with ions. $\boldsymbol{M}$ from (Fig. 10(c)) in the reverse bias state indicates Ohmic conduction due to a metallic cord between the conducting plates and the presence of a significant amount of uniformly confined charges. From (Fig. 10(d)), in the positive bias, an exponential to uniform homogenization effect of the capturing sites is evident. At -9.4 volts, a stronger and more intricate metallic connection takes place. Fig. 11 picturesises the first LRS state in the forward sweep of the chip when subjected to -$73^o$C. Region $\boldsymbol{O}$ is homologous to Region $\boldsymbol{L(i)}$ and $\boldsymbol{L(ii)}$ but with a higher slope. It suggests that increasing the temperature reduces the barrier height of the traps, thereby increasing the population of free carriers [34]. Rest all processes resemble the case of -$170^o$C.

### C. Conclusion

In this study, we employ the spin coating technique with PVDF-HFP and cuprous chloride composite to fabricate an

RSD. The device exhibits a WORM type of switching behavior at a relatively low operating voltage. This non-volatile characteristic can be attributed to the formation of a thick and stable copper filament within the polymer. We have observed that the reduction of CuCl to copper (Cu) atoms occurs through an electrochemical process that involves the application of a certain SET voltage. This process can be understood in the context of the electrochemical reduction potential of the copper(I) chloride (CuCl) system. A UV-Vis study has also been conducted on the device to provide additional insights into its working mechanism. The actual voltage required to reduce CuCl to Cu atoms in an electrochemical cell may depend on practical considerations such as cell design, solution concentration, temperature, and the presence of other ions or impurities. In our case, we have observed that CuCl forms a stable conducting channel of Cu atoms in the PVDF/HFP matrix at -8.7 volts. Since defects, temperature, and poling conditions vary, these values are also minutely affected in this case. The charge-transport phenomenon in the memristor under ambient, poling, and cooling conditions is also studied in much depth. Dominatingly TCLC, Ohmic, and complex ion hopping processes are found to be the underlying conduction mechanisms in the system. The field of organic electronics, particularly in devices like resistive switching devices or write-once-read-many devices based on materials like PVDF requires high voltages (approximately 20-60V) to achieve desired performance [16, 33]. This high voltage demand has recently been addressed by incorporating ionic liquids, reducing the operational requirement to less than 10V [9, 41]. However, our study proposes an alternative approach involving the utilization of copper (Cu) which holds the potential to render PVDF, a widely used commercial polymer, suitable for RSD applications at significantly lower operational voltages. The suggested study entails a comprehensive investigation into the influence of copper nanoparticles or electrodes with varying compliance, voltage, and concentration parameters. This exploration promises to shed light on the viability of employing copper to enhance the performance of PVDF-based organic RSDs while achieving improved operational efficiency and addressing voltage constraints. Such devices hold potential applications in various storage media, including Compact Disc-Recordable (CD-R), Digital Versatile Disc-Recordable (DVD-R), Blu-ray Disc Recordable (BD-R), and Write Once USB Drives [42].

### D. Conflict of interests

There is no conflict of interest.

### ACKNOWLEDGMENTS

We acknowledge NRF and CRF (IIT Delhi) for EBD, maskless-lithography, AFM and Probe Station facility. SB acknowledge MHRD, India for fellowship through IIT Delhi. SB also thanks Prof.Sameresh Das (IIT Delhi) for the experimental setup. . . . .